\documentclass[aps,twocolumn,floats,showpacs,showkeys]{revtex4}
\usepackage{graphicx}
\usepackage{epsfig}

\newcommand{\bastar}{\begin{eqnarray*}}
\newcommand{\eastar}{\end{eqnarray*}}
\newskip\humongous \humongous=0pt plus 1000pt minus 1000pt

\relax
\newcommand{\be}{\begin{equation}}
\newcommand{\ee}{\end{equation}}
\newcommand{\bea}{\begin{eqnarray}}
\newcommand{\eea}{\end{eqnarray}}
\newcommand{\pro}{\partial}
\newcommand{\dfrac}{\displaystyle\frac}
\newcommand{\ba}{\begin{array}}
\newcommand{\ea}{\end{array}}

\newcommand{\nn}{\nonumber}
\newcommand{\tF}{{\tilde F}}
\newcommand{\s}{\sum_{n=1}^{\infty}}
\newcommand{\ra}{\rightarrow}
\newcommand{\A}{\partial_a \delta {\cal L}}
\newcommand{\B}{\partial_b \delta {\cal L}}
\newcommand{\C}{\partial_a^2 \delta {\cal L}}
\newcommand{\D}{\partial_b^2 \delta {\cal L}}
\newcommand{\tA}{\partial_a {\cal L}}
\newcommand{\tB}{\partial_b {\cal L}}
\newcommand{\tC}{\partial_a^2 {\cal L}}
\newcommand{\tD}{\partial_b^2 {\cal L}}
\begin{document}
\title{Light Propagation Effects In QED: Effective Action Approach}
\bigskip
\author{Y. M. Cho}
\email{ymcho@yongmin.snu.ac.kr}
\affiliation{Department of Physics, College of Natural Sciences, Seoul National 
University, Seoul 151-742, Korea}
\author{D. G. Pak}
\email{dmipak@phya.snu.ac.kr}
\affiliation{Center for Theor. Physics, Seoul National University,
Seoul 151-742, Korea}
\affiliation{Institute  of  Applied
Physics,  Uzbekistan  National  University, 
Tashkent  700-095, Uzbekistan}
\author{M. L. Walker}
\email{m.walker@aip.org.au}
\affiliation{Department of Physics, Chiba University, 1-33 Yayoi-Cho, Inage-ku, Chiba 263-8522, Japan}

\begin{abstract}

The effects of light propagation in constant magnetic and electric
backgrounds are considered in the framework of the effective action
approach. We use the exact analytic series representation for the
one-loop effective action of QED and apply it to the birefringence effect.
Analytical results for the light velocity modes are obtained for weak and
strong field regime. We present asymptotic formulae for the light velocity
modes for the ultra strong magnetic field which can be realized in some
neutron stars. 
\end{abstract}
\vspace{0.3cm}
\maketitle 
\bigskip

\section{Introduction}

It is well known that the Euler-Heisenberg 
effective Lagrangian in quantum electrodynamics (QED) is presented by 
an asymptotic series \cite{euler}. 
Since the asymptotic series is divergent its
applications are limited to a weak field approximation. 
For the problems related to a strong field background 
one has to apply the complete expressions for the effective action. 
The effective action at one- and two-loop level
for special electromagnetic backgrounds  have been studied 
in \cite{ritus,ditt} (see as well \cite{dittbook} and refs. therein). 
Recently the analytic series representation for the
one-loop effective action of QED has been 
constructed \cite{qed1}
from Schwinger's integral expression for the effective action
\cite{schw}.  
The knowledge of an explicit analytical expression of the effective action
is useful for studying quantum effects like photon splitting \cite{adler}
and photon propagation in electric and magnetic fields 
\cite{tsai,shabad,dittgies,brezin}.
Another motivation comes from the theoretically predicted existence of 
strong magnetic fields of magnitude $B$ above the critical
value $B_{cr} =\dfrac{m^2}{e}
 \simeq 5 \cdot 10^{13} G$ in the cores of some
neutron stars \cite{pulsar}. 

In the present paper we consider light propagation
effects in the constant magnetic and electric field  
in the framework of the
effective action approach proposed in \cite{dittgies} 
applying the series representation for the one-loop QED effective action
 \cite{qed1}. In Section \ref{sec:lightcone} we present 
some general formulae relevant to the effective action approach \cite{dittgies}.
In sections \ref{sec:magnetic} and \ref{sec:electric} we consider 
photon propagation in the
pure magnetic and pure electric external field correspondingly. Birefringence effect
in the magnetic and electric field 
is considered in sections \ref{sec:birefringB}, \ref{sec:birefringE}. 
Some technical details of calculation and useful formulae 
are collected in the appendix.

\section{Light cone condition in the effective action formalism } \label{sec:lightcone}

The integral representation for the one-loop effective action
of QED for slowly varying background fields has been known for a long time
\cite{schw}
\bea
\Delta {\cal L} = - \dfrac{e^2}{8\pi^2} ab \int_{0}^{\infty} \dfrac{dt}{t}
\coth (eat) \cot (ebt) e^{-m^2 t}, \label{eq:schw}
\eea
where $a,b$ are gauge invariant variables corresponding to 
magnetic and electric fields in a Lorentz frame
where the both field vectors are parallel
\bea
&&a = \dfrac{1}{2} \sqrt {\sqrt {F^4 + (F \tF)^2} + F^2}, \nn \\
&&b = \dfrac{1}{2} \sqrt {\sqrt {F^4 + (F \tF)^2} - F^2}, \nn \\
&&F^2 = F_{\mu\nu} F^{\mu\nu}, \nn \\
&&\tF^{\mu\nu}=\dfrac{1}{2} \epsilon^{\mu \nu \rho\sigma} F_{\rho \sigma}. 
\eea

We start from the series representation for the
one-loop effective Lagrangian \cite{qed1} which
is more suitable in physical problems
when we need explicit analytical expression of the
effective Lagrangian
\bea
&{\cal L}&= -\dfrac{a^2 - b^2}{2} \nn \\
&-& \dfrac{e^2 }{4\pi^3} ab  \sum_{n=1}^{\infty}\dfrac{1}{n}
\Big [\coth(\dfrac{n \pi b}{a}) \Big ( {\rm ci}(\dfrac{n \pi m^2}{ea})
\cos(\dfrac{n \pi m^2}{ea}) \nn \\
&+&{\rm si}(\dfrac{n \pi m^2}{ea}) \sin(\dfrac{n \pi m^2}{ea})\Big )
- \dfrac{1}{2} \coth ( \dfrac{n \pi a}{b}) 
\Big ( \exp(\dfrac{n \pi m^2}{eb}) \nn \\
&\cdot& {\rm Ei}(-\dfrac{n \pi m^2}{eb} )  
+ \exp(-\dfrac{n \pi m^2}{eb} ) {\rm Ei}(\dfrac{n \pi m^2}{eb} )
\Big )\Big ] \nn \\
 &&\equiv {\cal L}_{cl}+\delta {\cal L}, \label{L0}
\eea
where ${\cal L}_{cl}$ is the classical QED Lagrangian.

We will follow the effective action approach \cite{dittgies}
to study the light propagation effects in non-trivial vacua.
The light cone condition 
for a soft photon propagating in the homogeneous electromagnetic
background (neglecting the back-reaction effects) can be derived 
from the equation of motion corresponding to the full effective action.
The final expression for the light cone condition is the following 
\cite{dittgies}
\bea
  0 &=& (\pro_x {\cal L} ) k^2  + M^{\mu \nu}_{\alpha \beta} k_\mu k^\alpha 
      \epsilon^\beta \epsilon_\nu, \nn \\
\bar k^\mu &=& \dfrac{k^\mu}{| {\bf k} |} \equiv (v , \hat {\bf k}),
 \label{lc}
\eea
where $\epsilon^\mu$ is a unit polarization vector of the 
photon, and $M^{\mu\nu}_{\alpha\beta}$ is defined by
(in the Lorenz gauge $k_\mu \epsilon^\mu=0$)
\bea
  M^{\mu \nu}_{\alpha \beta} 
      &\equiv& F^{\mu \nu } F_{\alpha \beta} (\pro^2_x {\cal L}) +   
 {\tilde F}^{\mu \nu }{\tilde F}_{\alpha \beta} (\pro^2_y {\cal L}) \nn \\
  && + \pro_{xy} {\cal L} ( F^{\mu \nu } {\tilde F}_{\alpha \beta} +
 {\tilde F}^{\mu \nu } F_{\alpha \beta} ), \nn \\
x &\equiv& \dfrac{1}{4} F_{\mu \nu} F^{\mu \nu} = \dfrac{1}{2}
 (a^2 - b^2), \nn \\
y &\equiv& \dfrac{1}{4}  F_{\mu \nu} {\tilde F}^{\mu \nu} = ab . 
\eea
The matrix $M^{\mu\nu}_{\alpha\beta}$ provides the 
quantum corrections to the classical light cone condition
$k^2=0$.
The light cone condition can be also transformed into another form
which is more suitable for 
calculation of the light velocity averaged over polarizations. 
After averaging over the polarization states of the photon the 
light cone condition can be rewritten as \cite{dittgies}
\bea
{\bar v}^2  &=& 1 - Q <T^{\mu \nu} > \bar k_\mu \bar k_\nu, \label{vbar}
\eea
 here, $ <T^{\mu \nu} >$ is a vacuum expectation value (VEV) of the 
electromagnetic energy-momentum  tensor, and $Q$ is an effective action charge 
defined in terms of the effective Lagrangian and its derivatives \cite{dittgies}.
For convenience we give the definition of the effective action charge $Q$ 
 in terms of gauge invariant variables $a, b$ 
\bea 
&  Q = (\pro^2_a  {\cal L} + \pro^2_b  {\cal L}  )  
       {\Big ( } {\cal L} (\pro^2_a  {\cal L}  + \pro^2_b {\cal L}) + \nn \\
& \dfrac{1}{a^2 + b^2 } (a \pro_a 
     {\cal L} - b \pro_b {\cal L})^2 - a \pro_a {\cal L} \pro^2_b {\cal L} 
      -  b \pro_b {\cal L} \pro^2_a {\cal L} {\Big )}^{-1} . \label{Qdef}
\eea
As can be seen from the light cone condition, Eq. (\ref{vbar}),
the light velocity is determined by the product
of the effective action charge  $Q$  and the  VEV of the energy-momentum tensor
$<T_{\mu\nu}>$.
It is convenient to introduce the following notations
for the combinations of special functions
\bea 
g(x) &=& {\rm ci}(x) \cos x + {\rm si}(x) \sin x , \nn \\
h(x) &=& \dfrac{1}{2} \Big [ e^x {\rm Ei} (-x) + e^{-x} {\rm Ei} (x) \Big ] .
\eea
The properties of the special functions are given in \cite{grad,stegun}.
Using Eq.~(\ref{L0})  for the one-loop correction $\delta {\cal L}$,
one can write down explicit expressions for the
derivative terms 
\bea
&\pro_a \delta {\cal L}&  = \dfrac{1}{a} \delta {\cal L}  - \dfrac {e^2 }{4 \pi^3}
    \sum^{\infty}_{n=1} \Big [ \dfrac {\pi b^2}{a} {\rm csch}^2 \beta g(z) \nn \\
&+& \pi a {\rm csch}^2   \alpha h(\bar z)- 
\dfrac {\pi m^2 b}{e a} g' (z) \coth \beta  \Big ], \nn \\
 &\pro_b \delta {\cal L} &=
 -\dfrac{e^2 }{4 \pi^3} \sum^{\infty}_{n=1} \Big [ \Big (\dfrac {a}{n} \coth \beta -
      \pi b {\rm csch}^2 \beta \Big ) g(z)  \nn \\
&-& \Big ( \dfrac{a}{n} \coth \alpha 
    + \dfrac{\pi a^2}{ b} {\rm csch}^2 \alpha \Big ) h(\bar z)  \nn \\
&+&
         \dfrac{\pi m^2 a}{ e b}  h' (\bar z) \coth \alpha
                              \Big ],  \nn \\
& \pro^2_a \delta {\cal L}& =  
 \dfrac{e^2 b }{4 \pi^3} \sum^{\infty}_{n=1} \Big [ \dfrac{1}{na} \coth \beta
            +
            g(z) \coth \beta\Big ( \dfrac { n \pi^2 m^4}{e^2 a^3}\nn \\
&-&      \dfrac{2 b^2 n \pi^2}{a^3} {\rm csch}^2 \beta \Big) 
    + \dfrac{2 b n\pi^2 m^2} {e a^3 } {\rm csch}^2 \beta g' (z) \nn \\
&+& \Big( \dfrac 
     {2 a n \pi^2 }{b^2} \coth \alpha -
              \dfrac{2 \pi}{b} \Big) {\rm csch}^2 \alpha h(\bar z) \Big ], \nn \\
&\pro^2_b \delta {\cal L}& =
\dfrac{e^2 a}{4 \pi^3} \sum_{n=1}^{\infty} \Big [ -\dfrac{1}{bn} \coth \alpha 
   +  h(\bar z)  \coth \alpha
  \Big ( \dfrac{n \pi^2 m^4}{e^2 b^3 }  \nn \\
&+& \dfrac{2 a^2 n \pi^2}{b^3 } {\rm csch}^2 \alpha
               \Big ) 
  - \dfrac{2 a n \pi^2 m^2 }{ b^3 e} {\rm csch}^2 \alpha h' (\bar z)   \nn \\
&+&     \Big ( \dfrac{2 \pi}{a} - \dfrac{ 2 b \pi^2 n }{ a^2} \coth \beta \Big)
     {\rm csch}^2 \beta g(z) {\Big ]},\label{QABCD}
\eea
where
\bea
  z &=& \dfrac{\pi n m^2}{ea}, ~~~~~\bar z = \dfrac{\pi n m^2}{e b}, \nn \\
\alpha &=& \dfrac {\pi n a}{b}, ~~~~~
\beta = \dfrac{\pi n b }{a} . 
\eea

The vacuum averaged value of the energy-momentum tensor
can be found from the effective Lagrangian as in \cite{dittgies}
\bea
<T^{\mu \nu}> &=&
-  \dfrac{T^{\mu \nu}_{cl} }{a^2 + b^2} \Big (
   a \tA - b \tB  \big ) + \nn \\
   && g^{\mu \nu } \Big [ {\cal L} - \dfrac{1}{2} (a \tA
 + b \tB) \Big ],
\label{Tmn}
\eea
here, $T^{\mu \nu}_{cl}$ is the classical energy-momentum tensor
\bea
{T^{\mu}_{cl}}_\nu = F^{\mu \rho} F_{\nu \rho} - \dfrac{a^2 - b^2 }{2} \delta^\mu_\nu.
\eea
%The vacuum expectation values of the Hamiltonian and   
%the trace of the energy-momentum tensor are
%\bea
%<T^{00}> &=& -{\cal L}  + b \tB, \nn \\
%<{T^\mu}_\mu> &=& 4 \Big [ {\cal L} - \dfrac{1}{2} \Big(a \tA + b \tB \Big ) \Big ]. 
%\eea
 In the following we will apply the above general equations 
to special cases of
light propagation in homogeneous pure magnetic and electric background.

\section{Light Velocity in Magnetic Field} \label{sec:magnetic}

Light propagation in the presence of a constant magnetic field in 
one-loop approximation
was first studied
in \cite{brezin,adler,tsai} and later by many others (see refs.~in \cite{dittbook}). 
The light velocity in the weak magnetic field was well 
studied using the Euler-Heisenberg effective
Lagrangian \cite{euler} (weak-field limit of the complete
one-loop effective QED Lagrangian)
\bea
&&{\cal L}_{E\!-\!H}={\cal L}_{cl}+\dfrac{2m^4}{\pi^2}\Big(\dfrac{ea}{m^2}\Big)^4\nn\\
&& \times {\sum_{n=0}^{\infty}}
  \dfrac{2^{2n} B_{2 n +4} }{(2n+2)(2n+3)(2n+4)} \Big (\dfrac{ea}{m^2}\Big)^{2n},
\label{EHlagr}
\eea
where $m$ is the electron mass and $B_n$ is the Bernoulli number. Clearly
the series (\ref{EHlagr}) is an asymptotic series which is divergent,
so that for ultra strong magnetic field problems one can not apply 
it for computational purposes. For instance, for the critical value $B_{cr}$
of the magnetic field,
which supposed to be in the cores of some
neutron stars, the series in the 
Euler-Heisenberg effective Lagrangian (\ref{EHlagr})
gives alternative in sign values when we include more than three
terms in the truncated series.

In this section we evaluate the light velocity 
using the series representation for the one-loop 
effective Lagrangian,
Eq.~(\ref{L0}), in weak and strong magnetic field regions and derive 
asymptotic formulae in the ultra strong field regime.

Without loss of generality we choose the magnetic field
to be directed along the $z$-axis, so that $\vec{B}  = (0, 0, a)$.
The non-vanishing  components of
the classical and quantum energy-momentum tensor are the following
\bea
T^{00}_{cl} &=& T^{11}_{cl} = T^{22}_{cl} =- T^{33}_{cl}
 = \dfrac{a^2}{2}, \nn \\
  <T^{00}> &=&- <T^{33}> = \dfrac{a^2}{2}     -  \delta {\cal L}_m  , \nn \\
 <T^{11}> &=& <T^{22}> = - <T^{33}>\nn \\
&& =\dfrac{a^2}{2} - 
\delta {\cal L}_m - a \A.
\eea
Hereafter in this section the functions $\A$, $\B$, $\C$, $\D$ are taken in the limit 
of vanishing electric field, $b \rightarrow 0$,
\bea
\A &=& -\dfrac{e^2 a}{2 \pi^4} G(a) - \dfrac{e^2 a^2}{4 \pi^4} G'(a) ,\nn \\
\B &=& 0, \nn \\
\C &=& -\dfrac{e^2}{2 \pi^4} G(a) - \dfrac{e^2 a}{\pi^4} G'(a) -\dfrac{e^2 a^2}{
4 \pi^4} G''(a), \nn \\
\D &=& \dfrac{e^4 a^2}{36 \pi^2 m^4} + \dfrac{e^4 a^3}{3 \pi^4 m^4} G'(a)\nn \\
&&+ \dfrac{e^4 a^4}{6 \pi^4 m^4} G''(a), \label{ABCDm}
\eea
where
\bea 
G(a) = &=& \s \dfrac{1}{n^2} g(\dfrac{n \pi m^2}{e a}) . \label{GGP}
\eea
The function $G(a)$ defines the one-loop contribution 
to the effective Lagrangian with a pure magnetic background 
and can be written in terms of the generalized Hurvitz $\zeta$-function.
From Eq.~(\ref{lc}) one can find immediately that the velocity of light
propagating  along the magnetic field remains  unchanged \cite{tsai,dittgies}
whereas for the photon moving in the plane orthogonal to the
magnetic field one has the  phase velocity given by 
\bea
 {\bar  v}^2 = \dfrac{1 - Q ( <T^{00}> + \dfrac{1}{2} <T^\mu_\mu>)}
            {1 + Q <T^{00}>}.
\eea
The VEVs of the energy-momentum tensor can be written in terms of the
main function $G(a)$
\bea
<T^{00}> &=& \dfrac{a^2}{2} - \delta {\cal L}_m ,   \nn \\
<T^\mu_\mu> &=&  \dfrac{e^2 a^3}{2 \pi^4} G'(a) ,  \nn \\
\delta {\cal L}_{m} &=& - \dfrac{e^2 a^2 }{4 \pi^4} 
                    G(a). \label{Qmag}
\eea
It should be noted that after taking the limit $b \rightarrow  0$ in the
general formulae (\ref{QABCD}) we have obtained a non-trivial contribution
from the  non-vanishing 
term $\pro^2_b {\cal L} {\Big |}_{b \rightarrow 0} $. This contribution
provides the leading decrease factor in the dependency of the 
light velocity in the strong field region. After 
simplification the final expression for the light velocity takes the form
\bea
{\bar v}^2  &=&\dfrac{2 a - \A - a \C}{2 a - \A + a \D}  \nn \\
 &=& {\Big (} 
       2 + \dfrac{e^2}{\pi^4}  G(a) + \dfrac{5 e^2 a}{4 \pi^4} G'(a)
+ \dfrac{e^2 a^2}{4 \pi^4} G''(a)
                                 {\Big )} \nn \\
&&
{\Big (} 
    2 + \dfrac{e^4 a^2}{36 \pi^2 m^4} + \dfrac{e^2 }{2 \pi^4} G(a) +
(\dfrac{e^2 a}{4 \pi^4} + \dfrac{e^4 a^3}{3 \pi^4 m^4}) G'(a) \nn \\
 &&+ \dfrac{e^4 a^4}{6 \pi^4 m^4} G''(a)
                          {\Big )}^{-1}.
\eea
To compare our results with the results obtained in past we
consider the light velocity
in weak field approximation which can be obtained 
using the truncated one-loop effective Euler-Heisenberg Lagrangian 
\cite{ritus,dittgies}
\bea
&&\tilde {\cal L}_{E-H} = {\cal L}_{cl} + \dfrac{\tilde c_1}{4} (b^2-a^2)^2
 +\tilde c_2 a^2 b^2, \nn \\
&& \tilde c_1 = \dfrac{8 \alpha_0^2}{45 m^2}, \nn \\
&&\tilde c_2=\dfrac{14 \alpha_0^2}{45 m^2} .
\label{eq:EH2}
\eea
 For the photon propagating in
the plane orthogonal to the magnetic field 
one can derive as in \cite{dittgies} 
the following expression for the velocity   
\bea
{\bar  v} = 1 - \dfrac {11\alpha^2_0} {45 m^4} B^2, 
\eea
here $\alpha_0 = \dfrac{e^2}{4 \pi \epsilon_0}=1/137.036$ is a fine structure constant.
The corresponding light velocities obtained from the exact one-loop 
effective Lagrangian (\ref{L0}) and one-loop Euler-Heisenberg 
Lagrangian (\ref{eq:EH2}) are plotted in Fig. 1.

\begin{figure}[t]
\begin{center}
\psfig{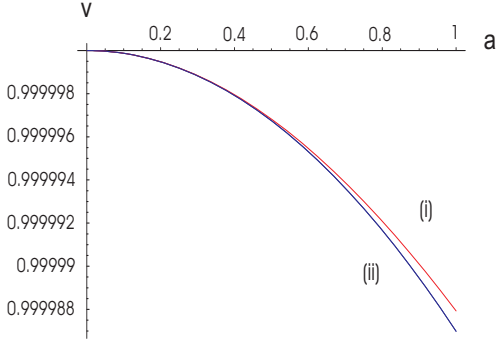}
\end{center}
\caption{\label{Fig. 1} Light velocity in the plane orthogonal to 
the magnetic field: (i)
exact one-loop approximated result; (ii) result in the weak field approximation.
The magnetic field strength $a$ is measured in units of electron mass $m^2$.}
\end{figure}

Now, let us consider light velocity in the strong field region.
From the asymptotic behaviour of the
main special functions ${\rm ci} (x), {\rm si} (x) $ 
\cite{grad,stegun} one can find the asymptotic formula for
the function $G(a)$
\bea
G(a) &=& -\dfrac{\pi^2}{6} ( \ln \dfrac {e a}{m^2} + c_1)
 -\dfrac{\pi^2 m^2}{2 e a} (\ln \dfrac{e a}{\pi m^2} +1)\nn \\
&&- \dfrac {\pi^2 m^4}{4 e^2 a^2} ( \ln \dfrac{2 e a}{\pi m^2} 
-\gamma+\dfrac{5}{2}), \nn \\
c_1 &=& -\gamma - \ln \pi +\dfrac{6}{\pi^2} \zeta'(2)=- 2.29191...  . \label{G1}
\eea

With this one results in the following 
asymptotic equation for the light velocity in a strong magnetic field
\bea
&&{\bar v}^2 \simeq \dfrac{ 1 - \dfrac{e^2}{12 \pi^2} ( \ln \dfrac {e a}{m^2} 
 + c_1 + 1) }{1 - \dfrac{e^2 }{ 12 \pi^2} ( \ln \dfrac{e a}{m^2} + c_2)
 + \dfrac{e^3 a}{ 24 \pi^2 m^2}} , \nn \\
&& c_2 = \ln {2}{\pi} + c_1 + \dfrac{3}{2} -\gamma
=-1.8207048...  \, . \label{vbarm}
\eea
The last term in the denominator in Eq.~(\ref{vbarm}) gives 
a linear decrease of light velocity
( for strong magnetic fields  
$a > B_{cr} = \dfrac{m^2}{e}$) in qualitative agreement with results
in \cite{tsai,dittgies}. 

Notice that the trace anomaly  $<{T^\mu}_\mu>$ 
of the energy-momentum plays a  crucial role in providing the
upper bound  for the factor $Q (<T^{00}> + \dfrac{1}{2} <{T^\mu}_\mu> ) 
   < 1  $
 until $a$ reaches the value of the Landau pole. 
Our formula improves on previously known expressions for the light velocity
\cite{tsai,dittgies} and can be
applied to ultra strong magnetic fields.

\section{Light velocity in homogeneous electric background} \label{sec:electric}

Let us consider an homogeneous electric field $\vec E = (0,0, b)$
directed along the $z$-axis. 
The non-vanishing components of the energy-momentum tensor and its VEVs
are
\bea
 T^{00} &=& T^{11} = T^{22} =- T^{33} = \dfrac{b^2}{2}, \nn \\
 <T^{00}> &=&- <T^{33}> = \dfrac{b^2}{2}  -  \delta {\cal L}_{el} + b \B , \nn \\
 <T^{11}> &=& <T^{22}>  = \dfrac{b^2}{2} + \delta {\cal L}_{el}  ,
\eea
As in the case of a pure magnetic background 
the  velocity of light propagating 
 along the electric field remains unchanged and 
in  the  directions orthogonal to the electric field we have for the light velocity
the same 
equation (16). The formula for the light velocity
derived from the one-loop Euler-Heisenberg effective Lagrangian
coincides with one expressed by Eq. (19) with the replacement $B \rightarrow E$.
The explicit expressions for the 
effective action charge $Q$  and VEVs of the components $<T^{00}>,<T^\mu_\mu>$
can be derived in a similar manner as in the previous section 
\bea
 Q &=& \dfrac{ \tC+\tD}{{\cal L}_{el} (\tC+\tD) + (\tB)^2 - b \tB \tC}, \nn \\
 {\cal L}_{el} &=& \dfrac{b^2}{2} + \dfrac{e^2 b^2}{ 4 \pi^4} 
      H(b) , \nn \\
 \A&=& 0 , \nn \\
 \B &=& \dfrac{e^2 b}{2 \pi^4} H(b) + \dfrac{e^2 b^2}{ 4 \pi^4} H'(b) , \nn \\
 \C&=& \dfrac{e^4 b^2}{36 \pi^2 m^4} + \dfrac{e^4 b^3}{3 \pi^4 m^4} H'(b)
+ \dfrac{e^4 b^4}{6 \pi^4 m^4} H''(b) , \nn \\
 \D&=&  \dfrac{e^2 }{2 \pi^4} H(b) + \dfrac{e^2 b}{\pi^4} H'(b)
  + \dfrac{e^2 b^2}{ 4 \pi^4} H''(b)   ,  \nn \\
  <T^{00}> &=& \dfrac{b^2}{2} + \dfrac{e^2 b^2}{4 \pi^4} H(b) + 
\dfrac{e^2 b^3}{4 \pi^4} H'(b) , \nn \\
 <{T^\mu}_\mu> &=& -\dfrac{e^2 b^3}{2 \pi^4} H'(b).
\eea
The function $H(b)$  is defined by
\bea
H(b) &=& \s \dfrac{1}{n^2} h(\dfrac{n \pi m^2}{e b}) . \label{HHF}
\eea

After simplifications we find the following
equation for light velocity in the orthogonal plane to the electric field
\bea
{\bar v}^2 &=& \dfrac{2 b  +\B - b \C}{2 b +\B + b \D} \nn \\
           &=& {\Big (}
 2  -\dfrac{e^4 b^2}{36 \pi^2 m^4} + \dfrac{e^2 }{2 \pi^4} H(b)
+ (\dfrac{e^2 b}{4 \pi^4} -\dfrac{e^4 b^3}{3 \pi^4 m^4})H'(b) 
\nn \\
 &&-\dfrac{e^4 b^4}{6 \pi^4 m^4} H'(b) 
                               {\Big )}
{\Big (} 2  + \dfrac{e^2 }{\pi^4} H(b) + \dfrac{5 e^2 b}{4 \pi^4} H'(b) \nn \\
 &&+ \dfrac{e^2 b^2}{4 \pi^4} H''(b)
                           {\Big )}^{-1}  .
\eea
    The exact values of the light velocity $\bar v$ 
in the weak field region $b \leq 1$ are plotted in Fig. 2 and can be compared with
ones obtained from the one-loop Euler-Heisenberg
Lagrangian (\ref{eq:EH2}).  
\begin{figure}[t]
\begin{center}
\psfig{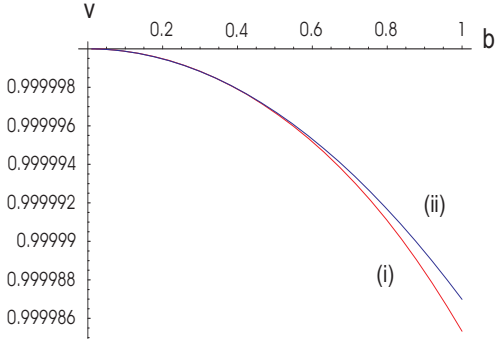}
\end{center}
\caption{\label{Fig. 2} Light velocity in the orthogonal plane to the 
electric field:
(i) the exact one-loop approximated result; 
(ii) the result obtained in weak field approximation.
The electric  field strength $b$ is given in units of electron mass $m^2$.}
\end{figure}

Of course, for a strong electric field of magnitude 
above the critical value
$E_{cr} = \dfrac{m^2}{e}$
pair creation, which we have not taken into account, becomes important, 
so in the strong field region one can only consider 
the light velocity dependence on the electric field $b$ formally.  
 In that case the light velocity dependence
 is determined by the asymptotic behaviour of the function $H(b)$
\bea
H(b) &=& 
          -\dfrac{\pi^2}{6} ( \ln \dfrac{e b}{m^2} + c_1) -\dfrac{\pi^3 m^2}
{4 e b} + \nn \\
&& \dfrac{\pi^2 m^4}{4 e^2 b^2} ( \ln \dfrac{2 e b}{ \pi m^2}
- \gamma + \dfrac{5}{2} ) , \label{HHFas}
\eea
where we keep only leading terms.

For a strong electric field one has the following asymptotic formula
\bea
&&{\bar v}^2 \simeq \dfrac{ 1 - \dfrac{e^2}{12 \pi^2} ( \ln \dfrac {e b}{m^2} 
 + c_2) }{1 - \dfrac{e^2 }{ 12 \pi^2} ( \ln \dfrac{e b}{m^2} + c_1+1)} .
\eea

In the asymptotic region approaching the Landau pole from the left
the light velocity squared
takes negative values. This shows that the Landau pole problem is still present.

\section{Birefringence in Magnetic Fields} \label{sec:birefringB}

In this section we consider
the birefringence effect with a complete one-loop
effective Lagrangian (\ref{L0}).
Let us 
choose the direction of the homogeneous magnetic field along the
$z$-axis and the wave vector $\hat {\vec k} $ in the $xOz$-plane
\bea
\bar {k}^\mu &=& (v, ~\sin \theta, ~ 0 , ~ \cos \theta ), \nn \\
\epsilon^\mu_\bot &=& ( 0, ~0, ~1, ~0), \nn \\
\epsilon^\mu_\parallel &=& (0, ~ -\cos \theta, ~0,~ \sin \theta) ,
\eea
where $\theta$ is the angle between the vectors $\vec {\bf B} , \hat {\bf k}$
and $\epsilon^\mu_\bot, ~
       \epsilon^\mu_\parallel $ are polarization vectors 
in orthogonal and parallel 
directions  to $\vec {\bf B}, \hat {\bf k}$-plane correspondingly.

To consider birefringence in the magnetic field we will
start from the light cone condition Eq. (\ref{lc}). 
The non-zero components of the matrix $M^{\alpha \beta}_{\mu \nu} $ are
\bea
M^{12}_{12} &=& a^2 \pro^2_x {\cal L} , \nn \\
M^{12}_{03} &=& a^2 \pro_{xy} {\cal L} , \nn \\
M^{03}_{03} &=& -a^2 \pro^2_y {\cal L} , \nn \\
M^{03}_{12} &=& -a^2 \pro_{xy} {\cal L}.
\eea

With this the light cone condition can be written as
follows
\bea
(v^2 - 1) \pro_x {\cal L} &=& a^2 \pro^2_y {\cal L} v^2 \epsilon_3 \epsilon_3
     + M^{mn}_{ab} {\hat k}_m  {\hat k}^a  \epsilon_n \epsilon^b +\nn \\
   &&  2M^{mn}_{03} v {\hat k}_m \epsilon_n \epsilon^3. 
\eea
For the $\bot, \parallel$-modes the last equation is simplified to
\bea
 (v^2_{\bot}-1) \pro_x {\cal L} &=& a^2 \sin^2 \theta \pro^2_x {\cal L}, \\
(v^2_{\parallel} -1) \pro_x {\cal L}
 &=& a^2 \sin^2 \theta \pro^2_y {\cal L} v^2_\parallel.
\eea
With this one can obtain the light velocities for 
$\bot,\parallel$ polarization modes 
\bea
v^2_\bot &=& 1 + \dfrac {a^2 \sin^2 \theta \pro^2_x  
{\cal L}}{\pro_x  {\cal L}}, \\
v^2_\parallel &=& \dfrac{1}{1 -
      \dfrac {a^2 \sin^2 \theta \pro^2_y  {\cal L}}{\pro_x  {\cal L}}},
\eea
or, equivalently, 
\bea 
v^2_\bot &=& 1 + a \sin^2 \theta \dfrac{\pro^2_a \delta {\cal L} - \frac{1}{a} 
\pro_a \delta {\cal L}}{\pro_a \delta {\cal L}} \, , \\
v^2_\parallel 
     &=& \dfrac{1}{1 + a \sin^2 \theta \dfrac{\pro^2_b \delta {\cal L} + \frac{1}{a} 
\pro_a \delta {\cal L}}{\pro_a \delta {\cal L}}} \,.
\eea
For the special case $\theta = \dfrac{\pi}{2}$
the light velocities for ($\bot , \parallel $)-modes
are plotted in Fig. 3.

\begin{figure}[t]
\begin{center}
\psfig{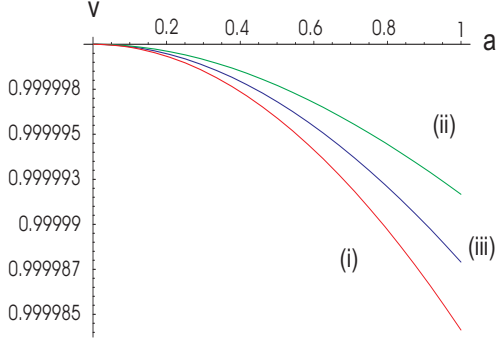}
\end{center}
\caption{\label{Fig. 3}
The upper curve, (ii), corresponds to the $\parallel$-mode and the lower
one, (i), corresponds to the $\bot$-mode. This can be 
compared with the light velocity
averaged over polarizations, curve (iii).}
\end{figure}

The estimated values for the velocities $v_{\bot, \parallel}$ 
can be compared with ones obtained in \cite{tsai}.
Tsai and Erber derived the explicit expressions
for the refraction indices for $\bot, \parallel$ modes
($\theta = \pi/2$)
\bea
n_{\bot, \parallel} = 1 + \dfrac{\alpha}{4 \pi} \eta_{\bot, \parallel}(h)
     + o(\alpha^2),
\eea
where $h = \dfrac{m^2}{2 e H}$ ($H$ coincides with our notation $a$)
and $\eta$ functions are expressed in terms of special functions.
This formula was derived approximately by neglecting
terms of order higher than $\alpha$.
So, the Tsai-Erber' formula is not valid for large magnetic fields
$a >> 1$. For weak fields our results are in a good quantative agreement with
Tsai-Erber' results ($\eta_{T-E}$) (see Table I).

\begin{table}[h]
\begin{center}
\begin{tabular}{|l|l|l|l|l|}
\hline
~h & ~~$\eta_{\parallel_{ T-E}}$ & ~~$\eta_{\parallel our}$ &
                   ~~$ \eta_{\bot_{T-E}}$ & ~~$\eta_{\bot our}$ \\
\hline
~1 & ~~0.0705  & ~~0.07019 & ~~ 0.0335 & ~~  0.03356  \\
\hline
~2 & ~~0.0189 & ~~0.01879  & ~~0.0101  & ~~ 0.01014\\
\hline
~3 & ~~0.00852 & ~~0.008484 & ~~0.00472  & ~~0.004725 \\
\hline
~4 & ~~0.00482  & ~~0.004801 & ~~ 0.00270  & ~~0.002708  \\
\hline
~5 & ~~0.00309 & ~~0.003082 & ~~0.00175 & ~~0.001748 \\
\hline
\end{tabular}
\caption{\label{tab:tc} Comparison of the Tsai and Erber results with
our results for the functions $\eta_{\parallel,\bot}$.}
\end{center}
\end{table}

For the strong field region $a>>1$ 
one can find the
asymptotic formulae  for the light velocities
\bea
{\bar v}^2_\bot &\simeq& \dfrac{ 1 - \dfrac{e^2}{12 \pi^2} ( \ln \dfrac {e a}{m^2} 
 + c_1+\dfrac{3}{2}) }{1 - \dfrac{e^2 }{ 12 \pi^2} ( \ln \dfrac{e a}{m^2} + c_1+
\dfrac{1}{2})} \,, \nn \\
{\bar v}^2_\parallel
 &\simeq& \dfrac{ 1 - \dfrac{e^2}{12 \pi^2} ( \ln \dfrac {e a}{m^2} 
 + c_1+\dfrac{1}{2}) }{1 - \dfrac{e^2 }{ 12 \pi^2} ( \ln \dfrac{2e a}{\pi m^2}
+1-\gamma) + \dfrac{e^3 a}{12 \pi^2 m^2}} \,. \label{strong}
\eea
In the asymptotic limit $a \rightarrow \infty$ we have
\bea
v^2_\bot  \rightarrow 1, ~~~~~~~~~~~~~
v^2_\parallel  \rightarrow - \dfrac{m^2}{ea} \ln (\dfrac{ea}{m^2})
 \rightarrow 0.
 \eea
One can also find the corresponding refraction indices
\bea
&& n^2_{\bot , \parallel} = \dfrac{1}{v^2_{\bot , \parallel}} .
\eea
For instance, keeping only the leading linear term in (\ref{strong}) one
can approximate the refraction index $n_\parallel$ by
\bea
&& n_\parallel \simeq \sqrt {1+\dfrac{e^3 a}{12 \pi^2 m^2} } \,\, ,
\eea
in agreement with other authors
\cite{shabad,shabadnew}. Notice also that the birefringence 
effect was considered recently in \cite{lor} 
where the effective action approach \cite{dittgies} 
had been applied to a one-loop part of the
Euler-Heisenberg effective Lagrangian (\ref{eq:EH2}).

\section{Birefringence in Electric Fields} \label{sec:birefringE}

Let us choose the direction of the electric field
along the $z$-axis, so that
\bea
 F^{\mu \nu} &=& 0 \,\,\,\,\, {\rm except}\,\,\, F^{03} = |{\bf {\vec E}}| \equiv b, \nn \\
 \tF^{\mu \nu} &=& 0 \,\,\,\,\, {\rm except}\,\,\, \tF^{12} = b .
\eea 
The photon momentum $\hat {\bf k}$ and polarization vector $\epsilon$ are the
same as ones considered for the pure magnetic case.
The non-zero components of the matrix $M^{\alpha \beta}_{\mu \nu} $ are
given by
\bea
M^{12}_{12} &=& b^2 \pro^2_y {\cal L} , \nn \\
M^{12}_{03} &=& -b^2 \pro_{xy} {\cal L} , \nn \\
M^{03}_{03} &=& -b^2 \pro^2_x {\cal L} , \nn \\
M^{03}_{12} &=& b^2 \pro_{xy} {\cal L}.
\eea
With this the light cone condition (\ref{lc}) can be rewritten as
follows
\bea
(v^2 - 1) \pro_x {\cal L} &=& b^2 \pro^2_x {\cal L} v^2 \epsilon_3 \epsilon_3
     + M^{mn}_{ab} {\hat k}_m  {\hat k}^a  \epsilon_n \epsilon^b \nn \\
&&-      2M^{03}_{ab} v {\hat k}^a \epsilon^b \epsilon_3. 
\eea
For  $\bot, \parallel$-modes the last equation gives 
\bea
 (v^2_{\bot}-1) \pro_x {\cal L} &=& b^2 \sin^2 \theta \pro^2_y {\cal L}, \\
(v^2_{\parallel} -1) \pro_x {\cal L} &=& b^2 \sin^2 \theta \pro^2_x {\cal L} v^2_\parallel.
\eea
After changing variables $(x,y) \ra (a,b)$ one obtains finally
\bea
v^2_\bot &=& 1 - \sin^2\theta  \Big ( \dfrac{b \pro^2_a \delta {\cal L} }
        {\pro_b \delta {\cal L}} +1  \Big ) , \nn \\
v^2_\parallel &=& \dfrac{1}{ 1 +  \sin^2\theta  
     \Big ( \dfrac{b \pro^2_b \delta {\cal L} }
        {\pro_b \delta  {\cal L}} -1  \Big )} , \nn \\
\eea
where derivatives of the $\delta {\cal L}$ are taken at the limit 
$a=0$.
For electric fields of magnitude below the critical value $b < 1$ one has 
a small deviation of the light velocity for $\bot, \parallel$-modes
from the velocity averaged over polarizations as it is pictured 
in Fig. 4. 

\begin{figure}[t]
\begin{center}
\psfig{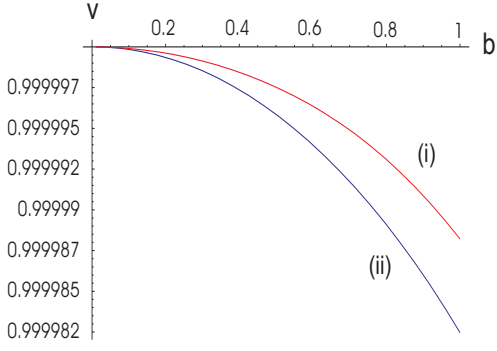}
\end{center}
\caption{\label{Fig. 4} Light velocity in the electric field
for the $\parallel $-mode, (i), and for the $\bot$-mode, (ii).}
\end{figure}

\section{Conclusion}

In this paper we have demonstrated the applications of the
analytic series representation for the one-loop effective action
of QED proposed recently \cite{qed1}. We have considered the light propagation
effects
in constant electric and magnetic fields
and demonstrated the advantage of our effective action approach 
in comparison with
the approach using the asymptotic series of Euler-Heisenberg 
effective Lagrangian 
(weak-field approximation of the complete one-loop effective Lagrangian)
in the case of strong and ultra-strong magnetic fields. 
The analytical asymptotic formulae for 
light speed and the refractive indices for arbitrarily strong magnetic fields
have been obtained.
Even though our analysis is restricted to one-loop
consideration, we hope that our results describe qualitatively the 
physical effects due to the very small value of the fine structure constant,
and can provide good quantitative results in astrophysical problems
concerning strong magnetic fields.
The generalization to two-loop order is of further interest.

{\bf APPENDIX} 

{\bf A. Notations and useful equations}

   Our conventions for the metric tensor $g_{\mu \nu}$ and antisymmetric tensor
$\epsilon^{\mu \nu \rho \sigma}$ are the following
\bea
g_{\mu \nu} &=& (\,-\,+\,+\,+\,) , \nn \\
\epsilon^{0123} &=& -1 , \nn \\
\epsilon^{123} &=& 1. 
\eea
We use Greek letters for 4-dimensional space-time coordinates and Latin letters
for space indices only.
   One has useful identities
\bea
 F_{\mu \rho} F_{\nu \sigma} F^{\rho \sigma} &=&  
\dfrac{1}{2} F^2 F_{\mu \nu} 
+ \dfrac{1}{2} (F \tilde F ) \tilde F_{\mu \nu} , \nn \\
F^{\mu \rho} F^\nu_\rho &-& \tF^{\mu \rho} \tF^\nu_\rho =
     2 x g^{\mu \nu}, \nn \\
F^{\mu \rho} \tF^\nu_\rho &=&  \tF^{\mu \rho} F^\nu_\rho =
     y g^{\mu \nu}.
\eea
Relations between the variables $a, b $ and $x, y$, and corresponding derivatives 
are given by
\bea
a &=& \dfrac{1}{2} \sqrt {\sqrt {F^4 + (F \tF)^2} + F^2}
    = \sqrt {\sqrt {x^2 + y^2  } + x } , \nn \\
b &=& \dfrac{1}{2} \sqrt {\sqrt {F^4 + (F \tF)^2} - F^2}
    = \sqrt {\sqrt {x^2 + y^2  } - x } , \nn \\
x &=& \dfrac{1}{4} F_{\mu \nu} F^{\mu \nu} = \dfrac{1}{2} (a^2 - b^2), \nn \\
y &=& \dfrac{1}{4}  F_{\mu \nu} {\tilde F}^{\mu \nu} = ab, \nn \\
\pro_x &=& \dfrac{1}{\rho^2} (a \pro_a - b \pro_b) , \nn \\
\pro_y &=& \dfrac{1}{\rho^2} ( b \pro_a + a \pro_b) , \nn \\
 \pro^2_x &=& \dfrac{1}{\rho^4} (a^2 \pro^2_a + b^2 \pro^2_b - 2 a b 
       \pro_{ab}) + \nn \\
    &&\dfrac{1}{\rho^6} [(3 ab^2 - a^3) \pro_a +
        (3 a^2 b - b^3) \pro_b], \nn \\
\pro^2_y &=& \dfrac{1}{\rho^4} (b^2 \pro^2_a + a^2 \pro^2_b + 2 a b 
       \pro_{ab}) - \nn \\
 &&\dfrac{1}{\rho^6} [(3 ab^2 - a^3) \pro_a -
        (3 a^2 b - b^3) \pro_b], \nn \\
\pro_{xy} &=& \dfrac{1}{\rho^4} (ab \pro^2_a - ab  \pro^2_b + (a^2 - b^2) 
       \pro_{ab}) - \nn \\
 &&\dfrac{1}{\rho^6} [(3 a^2b - b^3) \pro_a +
        (3 a b^2 - a^3) \pro_b], \nn \\
\rho^2 &=& a^2 + b^2.
\eea 

Using the change of variables $(x, y) \rightarrow (a,b)$  one can 
express  the matrix $M^{\mu\nu}_{\alpha \beta}$
in terms of variables $(a,b)$
\bea
&& M_{\alpha \beta}^{\mu \nu} 
 =\dfrac{1}{\rho^4} [ a^2 F^{\mu \nu} F_{\alpha \beta}  +
  b^2 \tF^{\mu \nu} \tF_{\alpha \beta} + ab (F^{\mu \nu} \tF_{\alpha \beta} \nn \\ 
 &+& \tF^{\mu \nu} F_{\alpha \beta})] \pro^2_a {\cal L} 
 +  \dfrac{1}{\rho^4} [ b^2 F^{\mu \nu} F_{\alpha \beta}  +
      a^2 \tF^{\mu \nu} \tF_{\alpha \beta}  \nn \\
 &-& ab (F^{\mu \nu} \tF_{\alpha \beta} + 
       \tF^{\mu \nu} F_{\alpha \beta})] \pro^2_b {\cal L} 
 + \dfrac{1}{\rho^4} [ -2 ab F^{\mu \nu} F_{\alpha \beta} \nn \\
 &+&   2 ab  \tF^{\mu \nu} \tF_{\alpha \beta} + (a^2 - b^2)  (F^{\mu \nu} \tF_{\alpha \beta} + 
               \tF^{\mu \nu} F_{\alpha \beta})] \pro_{ab} {\cal L} \nn \\
 &+&  \dfrac{1}{\rho^6} [ (3 ab^2 -a^3) F^{\mu \nu} F_{\alpha \beta}  -
      (3 ab^2 - a^3) \tF^{\mu \nu} \tF_{\alpha \beta} \nn \\
 &-&(3 a^2 b - b^3) (F^{\mu \nu} \tF_{\alpha \beta} + 
               \tF^{\mu \nu} F_{\alpha \beta})] \pro_a {\cal L} \nn \\
 &+&  \dfrac{1}{\rho^6} [ (3 a^2 b - b^3) F^{\mu \nu} F_{\alpha \beta}  -
      (3 a^2 b - b^3) \tF^{\mu \nu} \tF_{\alpha \beta}  \nn \\
&+&(3 ab^2 - a^3) (F^{\mu \nu} \tF_{\alpha \beta} + 
               \tF^{\mu \nu} F_{\alpha \beta})] \pro_b {\cal L}  . 
\eea

   For completeness we  write down the expressions for the energy-momentum
tensor $T_{\mu \nu}$ and effective charge $Q$ in terms of the $x, y$
variables \cite{dittgies}
\bea
&T^\mu_{\,\, \, \nu} = F^{\mu \rho} F_{\nu \rho} - x \delta^\mu_\nu, \nn \\
&<T^{\mu \nu} > = \dfrac{2}{\sqrt {-g}} \, \dfrac{\delta S}{\delta g_{\mu \nu}},
       \nn \\
&S= \int d^4x \sqrt{-g}  {\cal L}, \nn \\
&<T^{\mu \nu} > = - T^{\mu \nu} \pro_x {\cal L} + g^{\mu \nu} ( {\cal L} -
   x \pro_x {\cal L} - y \pro_y {\cal L} ), \nn \\
&Q= \dfrac{1}{2} (\pro^2_x {\cal L}+ \pro^2_y  {\cal L}) 
      {\Big (} (\pro_x {\cal L} )^2  + ( \pro_x {\cal L} ) ( \dfrac{x}{2} ( 
   \pro^2_x - \pro^2_y) \nn \\  
&+ y \pro_{xy} ){\cal L} + 
  \dfrac{1}{2} ( \pro^2_x + \pro^2_y) {\cal L} ( 1- x \pro_x - y \pro_y)
    {\cal L} {\Big )}^{-1}. 
\eea
 
{\bf B. Asymptotic formulae}

The asymptotic properties of standard  special functions are given in
\cite{stegun}. Keeping the leading terms in series expansions
we obtain the leading terms in asymptotic 
expansion of the main special functions: \\
 in  $x \rightarrow \infty$ region
\bea \label{eq:asymptotes}
 e^{-x} {\rm Ei} (x) & \ra& \dfrac{1}{x} ,\nn  \\
 e^x {\rm Ei}(-x) & \ra& - \dfrac{1}{x} , \nn \\ 
g(x) &\ra & - \dfrac{1}{x^2}  + \dfrac{6}{x^4} ,\nn \\
h(x) &\ra & \dfrac{1}{x^2}  + \dfrac{6}{x^4} ;
\eea
in $x \ra 0$ region
\bea
{ \rm Ei} (x) &\ra& \gamma + \ln x , \nn  \\
{\rm Ei}(-x) &\ra& \gamma + \ln x, \nn \\
{\rm ci} (x) &\ra & \gamma  + \ln x, \nn \\
{\rm si}(x) &\ra & -\dfrac{\pi}{2}, \nn \\
g(x) &\ra & \gamma + \ln x,\nn  \\
h(x) &\ra & \gamma + \ln x .
\eea
The asymptotic formulae for the 
functions $G(a)$, Eq. (\ref{G1}), and $H(b)$, Eq. (\ref{HHFas}),
can be derived in several ways, either from Schwinger's original
integral formula for the one-loop effective Lagrangian using the generalized
Hurwitz $\zeta$-function, or using the asymptotic formula for
$g(x)$ and the series representation Eq. (\ref{L0}).
Consider the integral representations for the main 
functions
\bea
g(x) &=& - \int_0^{\infty} \dfrac{t e^{-x t}}{1+t^2} dt , \nn \\
h(x) &=&  \int_0^{\infty} \dfrac{t e^{-x t}}{1-t^2} dt  .
\eea
Substituting these expressions  in Eqs. (\ref{GGP}, \ref{HHF}) and
performing the summation 
one finds the integral representations 
\bea
G(a) &=& - \int_0^{\infty} \dfrac{dt t }{1+t^2} 
          {\rm Li}_2(e^{-\frac{\pi m^2}{e a}t}) ,  \nn \\
H(b) &=&  \int_0^{\infty} \dfrac{dt t }{1-t^2} 
          {\rm Li}_2(e^{-\frac{\pi m^2}{e b}t})   \nn \\
\eea
from which the asymptotic formulae for the functions $G(a), H(b)$ 
can be easily derived. 
The values of the functions $G(a), H(b)$ obtained from
the series representation and from the asymptotic formulae
for large $a,b\gg 1$ can be viewed in Table II.
\begin{table}[h]
\begin{center}
\begin{tabular}{|l|l|l|l|l|l|}
\hline
~a & ~$G_{series}$ & ~$G_{asym} $ &
             ~b & ~$H_{series} $ & ~$H_{asym} $ \\
\hline
~$10^1$ & ~-0.37815  & ~-0.316198 &~$10^1$ & ~+ 0.122089 & ~+ 0.081690  \\
\hline
~$10^2$ & ~-2.38586 & ~-2.385448  &~$10^2$ & ~- 2.082565  & ~- 2.082954\\
\hline
~$10^3$ & ~-5.71865 & ~-5.718648 &~$10^3$& ~- 5.653113  & ~- 5.653116 \\
\hline
~$10^4$ & ~-9.42814  & ~-9.428142 &~$10^4$ & ~- 9.41787  & ~- 9.417870  \\
\hline
~$10^5$ & ~-13.2046 & ~-13.204571 &~$10^5$& ~-13.20317 & ~-13.203169 \\
\hline
~$10^6$ & ~-16.99072 & ~-16.990717 &~$10^6$ & ~-16.99054 & ~-16.9905396 \\
\hline
\end{tabular}
\caption{\label{tab2} Comparison of the results obtained from 
series representation with the results derived from the asymptotic formulae}
\end{center}
\end{table}

  {\bf Acknowledgments}

One of the authors (YMC) thanks 
Professor C. N. Yang for some illuminating discussions.
The work is supported in part by the
Korea Research Foundation and by
the BK21 project of the Ministry of Education. 
Some of this work while one of the authors (MLW) 
was working at the Department of Physics and Applied 
Physics at Kyung Hee Unversity.


\begin{thebibliography}{99}
\bibitem{euler} W. Heisenberg and H. Euler, Z. Phys. {\bf 98}, 714 (1936);
V. Weisskopf, Kgl. Danske Vid. Sel. Mat. Fys. Medd. {\bf 14},  6 (1936).
\bibitem {ritus} V. Ritus, JETP {\bf 42}, 774 (1976); {\bf 46}, 423 (1977);
M. Reuter, M. Schmidt, and C. Schubert, Ann. Phys, {\bf 259}, 313 (1997).
\bibitem  {ditt} A. Nikishov, JETP {\bf 30}, 660 (1970);
W. Dittrich, J. Phys. {\bf A9}, 1171 (1976);
S. Blau, M. Visser, and A. Wipf, Int. J. Mod. Phys.
{\bf A6}, 5409 (1991); J. S. Heyl and L. Hernquist, 
Phys. Rev. {\bf D55}, 2449 (1997);
C. Beneventano and E. Santangelo, J. Math. Phys. {\bf 42}, 3260 (2001).
\bibitem{dittbook} W. Dittrich, H. Gies, {\it Probing The Quantum Vacuum},
Springer Tracts in Modern Physics, Vol. 166; Springer, 2000.
\bibitem {qed1} Y. M. Cho and D. G. Pak, Phys. Rev. Lett. {\bf 86}, 1947 (2001);
W. Mielniczuk, J. Phys., {\bf A15}, 2905 (1982). 
\bibitem{schw}  J. Schwinger, Phys. Rev. {\bf 82}, 664 (1951).
\bibitem{adler} S. L. Adler, Ann. Phys. (N. Y.) {\bf 67}, 599 (1971).
\bibitem{brezin} Z. Bialynicka-Birula and I. Bialynucki-Birula, Phys. Rev.
{\bf D2}, 2341 (1970); E. Brezin and C. Itzykson, Phys. Rev. D {\bf 3}, 618 (1971).
\bibitem {tsai} W.-Y. Tsai and T. Erber, Phys. Rev. D {\bf 12}, 1132 (1975);
W.-Y. Tsai, Phys. Rev. {\bf D 10}, 2699 (1974); V. N. Baier, V. M. Katkov and
V. M. Strakhovenko, ZhETF {\bf 68}, 403 (1975).
\bibitem{shabad} I. A. Batalin and A. E. Shabad, ZhETF {\bf 60},
894 (1971) ({\it Sov. Phys.} - JETP {\bf 33},483 (1971).
\bibitem {dittgies} W. Dittrich and H. Gies, Phys. Rev. D{\bf 58}, 025004 (1998);
      Phys. Lett. B {\bf 431}, 420 (1998).
\bibitem{pulsar} V. M. Kaspi, {\it Pulsar Astronomy - 2000 and Beyond. ASP 
Conference Series} {\bf 202} (1999) (M. Kramer, N. Wex and R. Wielebinski,
eds.), astro-ph/9912284.
\bibitem {grad} I. S. Gradshtein and I. M. Ryzhik, {\it Tables of integrals,
  Series and Products}, Academic Press (1965).
\bibitem {stegun} M. Abramowitz and I. A. Stegun, {\it Handbook of Mathematical
Functions}, Dover Publ., Inc., NY (1972).
\bibitem{shabadnew} A. E. Shabad, {\it Interaction of Electromagnetic Radiation 
with Supercritical Magnetic Field}, hep-th/0307214.
\bibitem{lor} V. A. De Lorenci, R. Klippert, M. Novello and J. M. Salim,
Pjys. Lett. B {\bf 482}, 134 (2000). 
\end{thebibliography}
\end{document}